\documentstyle[aps,preprint]{revtex}
\tightenlines
\begin{document}
\title{ {\bf Ballistic electron transport in stubbed quantum waveguides:
 experiment and theory}}

\author{  P. Debray\cite{phil}$^1$, O. E. Raichev\cite{oleg},
P. Vasilopoulos\cite{takis},   M. Rahman$^2$,\\ R. Perrin$^1$,
and W. C. Mitchell$^1$\\
\ \\}

\address{$^1$ Air Force Research Laboratory, AFRL/MLPO, Wright-Patterson
AFB, Ohio, USA\\
\cite{phil}Service de Physique de l' Etat Condense, CEA de Saclay,
91191 Gif-sur-Yvette,
France\\
\cite{oleg}Institute of Semiconductor Physics, NAS,
45 Prospekt Nauki, 252650 Kiev,  Ukraine\\
\cite{takis}Concordia University, Department of Physics,
Montr\'{e}al, Qu\'{e}bec, Canada, H3G 1M8\\
$^2$Department of Physics, 
University of Glasgow, Glasgow, G12 8QQ, UK} 
\date{\today}
\address{}
\address{\mbox{}}
\address{\parbox{14cm}{\rm \mbox{}\mbox{}\mbox{}
We present results of experimental and theoretical investigations
of electron transport through stub-shaped waveguides or electron stub
tuners (ESTs) in the ballistic regime. Measurements of the
conductance $G$ as a function of voltages, applied to different
gates $V_i$ ($i=$bottom, top, and side) of the device, show oscillations
in the region of the first quantized plateau 
which we attribute to reflection resonances.
The oscillations are rather
regular and almost periodic when the height $h$ of the EST cavity
is small compared to its width. When $h$ is increased, the oscillations
become less regular and broad depressions in $G$ appear. A theoretical
analysis, which accounts for 
the electrostatic potential formed by the gates in the cavity region,
and a numerical computation of the transmission probabilities
successfully explains the experimental observations. An important
finding  for real devices, defined by surface Schottky gates,
is that the resonance minima result from
size quantization  along the transport direction of the EST.}}

\address{\mbox{}}
\address{\parbox{14cm}{ \rm PACS: 73.61.-r, 85.30.Vw}}
\maketitle

\section{INTRODUCTION}

Submicron-size T-shaped electron waveguides, defined electrostatically
in a two-dimensional electron gas (2DEG) by Schottky gates, are
very promising devices for potential applications in microelectronics since
their conductance $G$ is determined, in the ballistic regime, by quantum
interference effects and can be changed by applying voltages
to the gates \cite{1}. Such devices, commonly known as Electron Stub
Tuners (ESTs), also open the way for studying resonant states of
ballistic quantum dots in both the weakly coupled tunneling and in
the transmissive open regime \cite{2}. 
The size of an EST  can be controlled by gate voltages, cf. Fig. 1.
For a theoretical analysis, an EST can be considered as a rectangular
quantum dot connected to 2DEG reservoirs through two oppositely placed
Quantum Point Contacts (QPC). 
When the electron phase-coherence length exceeds the dimensions of
the EST, transport through the device is ballistic. A number of theoretical
papers \cite{3}, \cite{4} have been published on the ballistic transport
characteristics of ESTs in the open regime of the QPCs. These works predict
an oscillatory dependence of $G$ as a function of geometrical size
parameters of the device or of the Fermi energy $E_F$. A minimum in $G$,
or a reflection resonance, is said to occur due to resonant reflection
of electron waves by quasibound states of the stub cavity (SC) formed
by the gates, the quasibound state itself resulting because of the
quantization of electron momentum associated with the small device
size. Experimentally, it is possible to probe these resonance states
through measurements of $G$ at low temperatures and
has been reported \cite{4} only very recently.  
However, so far experimenters have failed to observe well-defined,
regular oscillations in $G$ with mimima corresponding to excitations
of the quasibound states as predicted theoretically. Most of the
devices so far used in experiments were geometrically defined by
only two gates, which do not allow an adequate independent control
of the width of the QPCs and the shape of the SC. This possibly
explains the failure to observe experimentally a well-defined,
regular pattern of minima in $G$.

In this paper we present experimental and theoretical results for the
four-gate EST. A preliminary account of some of them has appeared recently 
\cite{4a}. We report the experimental observation of a clear and pronounced
oscillatory dependence of the ballistic $G$ on the size of the SC as the latter
is changed by voltage-biasing the gates. Such oscillations occur on the first
conductance plateau of the QPCs. We also present  theoretical results
for $G$ obtained from a numerical solution of  Schr\"oedinger's
equation for a two-dimensional (2D) hard-wall electron waveguide with a shape
close to the one resulting from the biasing of the Schottky gates. Because of
this choice, we believe our results are closer to reality than those reported
in previous theoretical work based on a rectangular approximation for the SC
shape. A rectangular SC shape is unrealistic since,
though the lithographically defined device shape is rectangular, the shape of
the SC changes as the gates are biased \cite{5}. Moreover, except for  a
few papers \cite{6}, the lengths of the QPCs have been considered as
infinite; this is a very rough approximation for real devices and we avoid
it in our computations. Comparison of the experimentally observed features of
the ballistic $G$ to those obtained numerically enables us to
determine the physical origin of these features and helps us understand what
the shape of the SC is and how it can be modified by applied gate voltages.

The paper is organized as follows. In Sec. II we give a brief
description of the device fabrication and measurement techniques and then
present results of conductance measurements as function of gate voltages.
Section III outlines the theoretical model and 
of the calculations presents numerical results. Finally, an interpretation of 
the experimental results based on the theoretical analysis of Sec. III. and 
conclusions follow in Sec. IV.

\section{EXPERIMENTAL ASPECTS}

\subsection{Device fabrication and measurement techniques}

The ESTs used in this study were fabricated from modulation doped
AlGaAs/GaAs heterostructure wafers grown by MBE and having a
two-dimensional electron gas (2DEG) at a depth of 80 nm below the
surface. The carrier concentration $n_{2D}$ of the 2DEG was $2.4 \times
10^{15}$ m$^{-2}$ with a mobility $\mu$ of 100 m$^2$/V s at 4.2 K.
These values of $n_{2D}$   and $\mu$ give a 2DEG Fermi energy $E_F=$8 meV.
The ESTs were defined by four Schottky gates S1, T, S2, and B patterned
by electron beam lithography on the surface of the wafer. Figure 1
gives a schematic drawing of an EST device while Fig. 2 shows a scanning
electron micrograph of a fabricated EST sample. Lighter areas are the
Schottky metal gates on the wafer surface. The central part of the
EST, where the SC is  
located, forms a lithographically rectangular planar quantum dot
of length 0.55 and width 0.25 $\mu$m. The lithographic lengths of the QPCs
is 0.1 $\mu$m. The samples were clamped to the mixing chamber of a
dilution refrigerator. Considerable care was taken to ensure good
thermal contact to the sample. The two-terminal conductance G of the
devices was measured at 90 mK as function of a gate voltage, while
the other gates were biased at fixed voltages. Standard low-bias,
low-frequency, lock-in techniques were used to measure G, which was
corrected for a low series resistance due to the 2DEG reservoirs.
A source-drain rms excitation of 10 $\mu$V was typically used to drive
a current along the length ($x$-direction) of the QPCs.

Since the four gates are independent, it was possible to characterize
the QPCs of the EST device independently by biasing the appropriate gates while
grounding the rest of them. High-quality, well-defined conductance
quantization staircase was observed for both QPCs. The gates of the
QPCs were negatively  biased  
to assure fundamental-mode transport through them.
The conductance G of the device was then measured as function of the
size of the SC by sweeping the voltage $V_T$ of the top gate, or $V_B$
of the bottom gate, or $V_S$ of the side gates, while the other gates
were biased at fixed negative voltages. The
sweeping gate voltage was changed until the device was completely
pinched off, allowing measurements in both the single-mode open and
the tunneling regime of electron transport. Measurements were made
on a few EST devices. All of them gave nearly identical and
reproducible results differing only in the pinch-off voltages.

\subsection{Experimental results}

Figures 3 to 7 show the experimental results. Very well-defined
oscillations in the ballistic conductance G are observed as a function
of a sweeping gate voltage, which changes the size of the SC, while
the other gates are biased at fixed voltages. These oscillations
exhibit several features which are found to be generic to the EST
devices studied. All  results shown correspond to transport in
the fundamental mode through the QPCs until they are pinched off.

Figure 3 shows the oscillations in G observed when the bias voltage
$V_T$ of the top gate is swept, while the other gates are kept at
fixed voltages. The solid curve is obtained when the bottom and the
two side gates are biased, respectively, at $V_B= -765$ mV,
$V_{S1} = -860$ mV, and $V_{S2} = -964$ mV. As $V_T$ is made more
negative the size of the SC shrinks. This also adds to the depletion
due to the side gates and narrows  the QPCs until they are
pinched off when the device conductance drops to zero. The oscillations
in $G$ are found to occur in two distinct regimes of the sweeping gate
voltage, one for which $G < e^2/h$ and the other for which $G > e^2/h$.
The one with $G < e^2/h$ is the tunneling regime when $E_F$ is below the
bottom of the lowest conduction subband of the QPCs, which now form
energy barriers through which electrons can tunnel. The oscillations
of G in this regime are found to be periodic, quite sharp, and well
resolved. The regime for which $G > e^2/h$  may be called
the open regime. This happens when $E_F$  is above the bottom of
the lowest subband of the QPCs such that transport is in the fundamental
mode. The $G$ oscillations in this regime are located on the first
quantized conductance plateau of the QPCs. Though
quite robust, clear, and nearly periodic, they are relatively broad
and show a certain degree of overlapping with the resulting convolution
effect. The peak values are less than the expected quantized value
of $2e^2/h$ and, as will be seen later, result from backscattering
at the QPC entrance and/or from boundary roughness at the QPC
walls. The dotted $G$ curve of Fig. 3 is obtained for $V_B = -780$ mV
and differs from the solid one in two important respects. First, the
average conductance is substantially lower. Second, the oscillations
in $G$ in the open regime become irregular due to the appearance of
broad troughs or depressions in the conductance.

The size of the central ballistic cavity of the device can also be
altered by varying the bias voltages of the side gates while keeping
those of the top and bottom gates at fixed values.
Figure 4 shows the variation in $G$ as function of the side gate
bias voltage $V_S$ ($V_S = V_{S1} =  V_{S2} + 104$ mV). 
The solid curve was generated with $V_T = -1400$ mV and $V_B = -755$ mV, 
while the dotted curve was obtained for $V_T = -800$ mV and $V_B = -790$ mV. 
Results obtained by sweeping
$V_B$ are illustrated in Fig. 5. For these measurements the side
gates were biased as follows: $V_{S1}  = -860$ and $V_{S2} = -964$ mV.
The solid and dotted curves correspond, respectively, to $V_T= -1085$
and $-635$ mV. Comparing the results of Figs. 4 and 5 to those of
Fig. 3, we notice that, except for the device pinch-off voltages
and the oscillation periods, the
features of the oscillations in $G$ are similar. The solid curves show
the same characteristics, as do the dotted ones, though the features
are different for the two sets. This is not surprising since in all
cases we are changing the size of the central cavity. An interesting
question, however, is what causes the difference in the characteristic
features of the $G$ oscillations observed on the first conductance
plateau for the solid and the dotted curves. If we look more closely
and compare the constant gate bias voltages used for generating the
two sets of $G$ curves, an empirical consistency emerges. The voltages
used for the dotted curves are such as to result in a SC which is long
compared to that for the corresponding solid curves. As an example,
the oscillatory $G$ of the dotted curve in Fig. 4 is obtained
for a $V_T = -800$ mV, while for the solid curve $V_T$ is equal
to $-1400$ mV. A more negative top gate voltage certainly makes
the SC shorter. These observations lead us to conclude that a lower
average value of $G$ and broad depressions in it occur when the stub
cavity is long. We call the oscillatory $G$ pattern with regular
minima observed for the solid curves a "regular" pattern and that
for the dotted curves an "irregular" pattern.

In order to better understand the origin of the observed oscillations
in $G$ and to distinguish between the peaks in the tunneling and the
open regime, we have studied the dependence of the regular $G$
pattern on temperature, drain-source excitation voltage, and a
magnetic field applied perpendicular to the plane of the 2DEG. Figure
6 shows the temperature dependence of a regular $G$ pattern obtained
by sweeping the top gate voltage $V_T$ with $V_B = -765$ mV, $V_{S1}
= -860$ mV, and $V_{S2}  = -964$ mV. As the temperature is increased,
all peaks in both the tunneling and the open regime broaden, and
eventually they are washed out. At 4.2 K, the oscillatory $G$ pattern
has disappeared and is replaced by the conductance step and plateau.
At the highest temperature of 2.5 K measured in the dilution refrigerator,
the peaks in the open regime have practically disappeared, while those
in the tunneling regime show a trace existence. The influence of the
source-drain voltage $V_{ds}$ on the regular $G$ pattern has also been
studied and is shown in Fig. 7. Notice that the effect of increasing
$V_{ds}$ is similar to that of temperature. The oscillations in $G$
are found to practically fade out and be replaced by the conductance
step and plateau when $V_{ds}$ is increased to a rms value of 700 $\mu$V.

\section{THEORETICAL TREATMENT}
\subsection{Cavity potential}

A realistic, theoretical description of ballistic electron transport
through a cavity, such as the stub of an EST, must take into account
the electrostatic potential inside the cavity since it determines
the actual shape of the conducting channel. Accordingly, we have 
calculated the electrostatic potential created in the plane of a 2DEG 
situated at a distance $d=$80 nm below the surface, at $z=0$, of a
two-gate EST, defined by two surface Schottky gates with voltages
$V_T$ and $V_B$, when $S1$, $T$, and $S2$ are connected together, cf. Fig. 1. 
The distance between the gates at entrance and exit
is $2w=$250 nm, the bottom gate $G_B$ is flat, while the top gate
$G_T$ contains the stub-like opening of width $2w$ and of length
$L=$300 nm. This value of $d$ and the lithographic dimensions
correspond to the experimental device described above, although
the present model of just two gates is somewhat idealized but
necessary for simplifying the calculations. The potential
$\varphi(x,y,z)$ has been calculated from the Laplace equation
in the semi-space $z>0$, with Dirihlet boundary conditions
$\varphi(x,y,0)=V_i$ on the $i-$th gate region ($i=T,B$) and Newmann
boundary conditions $\partial \varphi(x,y,z)/\partial z |_{z=0}=
4 \pi e n_{2D}/\epsilon$ at the exposed surface region; $n_{2D}$
is the electron concentration in the 2D gas in the absence of depletion
and $\epsilon$ is the dielectric constant. The last boundary
condition expresses the so-called "frozen surface model" in which
the electric charge at the exposed surface is constant; this model
appears appropriate at low temperatures and is often used in
theoretical calculations\cite{8}. To make our model finite in
the $x$ direction, we choose a length $l=L$ and use the
boundary condition $\partial \varphi(x,y,z)/\partial x |_{x=\pm l}=0$.
We have also assumed that the concentration of the ionized donors
in the doping region between the surface and the 2D gas plane
is equal to a sum of the $n_{2D}$ and surface charge concentration and
is not changed appreciably when the voltages are applied to the gates.

In Fig. 8 we present the resulting contour plot of the potential
$\varphi(x,y,d)$; since  $\varphi(-x,y,z)=\varphi(x,y,z)$, 
we show $\varphi$ only for the right half of the stub.
Although we do not take into account the free-electron charge
in the plane $z=d$, in order to avoid the heavily involved
self-consistent calculations, we expect that the screening
effect due to these electrons will not change the shape of the
equipotential lines considerably but would cause at most a flattening
of the bottom of the potential distribition.
As a result, we expect the shape of the conducting channel
to follow, more or less, the calculated equipotential lines. This
enables us to draw the following important qualitative conclusions.

i) The shape of the cavity inside the stub region does not follow
that defined by the gate edges and is not rectangular as  has been assumed
in previous pertinent theoretical works.\\ 
ii) The width of the cavity is close to the lithographic
width of the stub, and since the Fermi wavelength at $E_F
\simeq$ 8 meV is about 53 nm, which is considerably less than the
lithographic stub width $2w$, the cavity accommodates not just one
longitudinal mode, as has been frequently assumed, but several modes.\\
iii) When the width of
the narrowest part of the conducting channel, in our model at $x=l$, is
small compared to the lithographic one $2w$ of the wire, the length
of the cavity at $x=0$ is considerably smaller than the lithographic
length $L$.\\
iv) The length of the cavity is even smaller when the
upper gate voltage $V_T$ is more negative ($V_T<V_B$) so that there is
an overall shift of the conducting channel towards the bottom gate.
In going beyond the two-gate model towards the four-gate device shown in
Fig. 1, it is reasonable to expect that when the top gate voltage is more
negative than that of the side gates, the height of the cavity
decreases, while in the opposite case it should increase. \\
In the following we use this qualitative information
to appropriately model the shape of the conducting channel in
the four-gate EST and calculate the electron transmission through the cavity.

\subsection{Model of the cavity and numerical method}

We model the conducting channel of the device in Fig. 1 with a
2D waveguide having 
hard-wall boundaries described, in an obvious notation, by the functions

\begin{eqnarray}
y_{bot}(x)=-y_{wire}(x),~~~ y_{wire}(x)=W/[1+\exp((-x+r)/\beta)]
+W/[1+\exp((x+r)/\beta)], \\
y_{top}(x)= y_{wire}(x) + a + y_{cav}(x),~~~y_{cav}(x)=h \exp(-x^2/b^2).
\end{eqnarray} 
We describe the cavity with the Gaussian function $y_{cav}(x)$
since it gives us the most relevant elementary-function approximation of
the equipotential lines shown in Fig. 8. The function $y_{wire}(x)$
describes the transition from the constricted region near $x=0$ to the 2D
reservoirs at $x=\pm \infty$. Here we set $\beta=W/4$ to model the
square-angle opening of the conducting channel of the experimental
device (Fig. 1). The parameter $W$, which describes the semiwidth of the
channels away from the constriction and must be large enough, is chosen
as $W=w$. For this value of $W$, the channels away from the constriction
already accommodate about 10 transverse modes and can be effectively
treated as 2D leads. The remaining parameter $r$ is chosen, by inspection,
as $r=w+l+3 \beta$, where $l$ is the lithographic length of the QPCs;
this gives a more or less suitable correspondence between the outer
parts of the conducting channel and the gate corners. The resulting
shape of the conducting channel, together with the lithographic gate
layout, is shown in Fig. 9.

To determine the transmission coefficients of electron waves
through the device, we solved numerically the 2D Schr{\"o}dinger equation

\begin{equation}
-\frac{\hbar^2}{2m} \left( \frac{\partial^2}{\partial x^2} +
\frac{\partial^2}{\partial y^2} \right)\Psi(x,y) +
[U(x,y) - \varepsilon ] \Psi(x,y)=0,
\end{equation}
using the following expansion for the wave function \cite{9}
\begin{equation}
\Psi(x,y) = \sum_{n} \psi_n(x) \chi_n(x,y),~~~\chi_n(x,y)=
\sqrt{\frac{2}{Y(x)}} \sin \left[\frac{\pi n}{Y(x)}(y-y_{bot}(x))
\right],
\end{equation}
where $Y(x)=y_{top}(x)-y_{bot}(x)$ is the $x$-{\it dependent} channel width.
The basis functions $\chi_n(x,y)$ already satisfy
the boundary conditions for  hard-wall confinement. Substituting
Eq. (4) into Eq. (3)  leads to the 1D matrix equation for $\psi_n(x)$

\begin{equation}
\left[ \frac{d^2}{d x^2} - \left(\frac{\pi n}{Y(x)} \right)^2
+ k^2 \right] \psi_n(x) + \sum_{m}
\left[ 2B_{nm}(x) \frac{d}{d x}
+ C_{nm}(x) - K_{nm}(x) \right] \psi_{m}(x) =0;
\end{equation}
here $k^2=2m \varepsilon /\hbar^2$ and

\begin{eqnarray}
B_{nm}(x)=
\int_{\textstyle y_{bot}(x)}^{\textstyle y_{top}(x)}
dy \chi_n(x,y) \frac{\partial}{\partial x} \chi_m(x,y), \nonumber \\
C_{nm}(x)=
\int_{\textstyle y_{bot}(x)}^{\textstyle y_{top}(x)}
dy \chi_n(x,y) \frac{\partial^2}{\partial x^2} \chi_m(x,y), \\
K_{nm}(x)= \frac{2m}{\hbar^2}
\int_{\textstyle y_{bot}(x)}^{\textstyle y_{top}(x)}
dy \chi_n(x,y) U(x,y) \chi_m(x,y). \nonumber
\end{eqnarray}
Since we assume $U(x,y)=0$ far away from the constriction, all
parameters defined by Eqs. (6) depend on $x$ only in the constriction
region. We choose $x_{max}$ and $x_{min}$ far enough from the
constriction and discretize Eq. (5) on a $N+1$-point grid according
to $x=x_i=x_{min}+i s$, $s=(x_{max}-x_{min})/N$. The resulting
finite-difference equation for $\psi_{m}(x_i)$ 
is solved subject to the boundary conditions
$A_{nm}(1)\psi_{m}(x_1)+A_{nm}(0)\psi_{m}(x_0)=A^{\alpha}_n$ and
$A_{nm}(N-1)\psi_{m}(x_{N-1})+A_{nm}(N)\psi_{m}(x_N)=0$, appropriate to 
a wave, in state $\alpha$, incident from the left side.
Since $\chi_n(x,y)$ are the exact normalized
eigenfunctions of the problem at $x=\pm \infty$, the
boundary matrices $A_{nm}$ are diagonal 
$A_{nm}(1)=A_{nm}(N-1)=\delta_{nm}, 
A_{nm}(0)=A_{nm}(N)=-\delta_{nm} \exp(-i p_n s)$, while
$A^{\alpha}_n=\delta_{n \alpha}[\exp(i p_n s)-\exp(-i p_n s)]$. 
In these matrix expressions we introduced the longitudinal quantum number

\begin{equation}
p_n=\sqrt{ k^2-(\pi n/Y(\infty))^2},
\end{equation}
which can be either real or imaginary ($Im~p_n>$0);
in the latter case the waves are evanescent in the leads.
Far away from the constriction $p_n$ is the longitudinal momentum in the leads.

The ballistic conductance $G$ at zero temperature is given by
the multichannel Landauer-B{\"u}ttiker formula

\begin{equation}
G=\frac{2e^2}{h} \sum_{\alpha \alpha'} |T_{\alpha \alpha'}|^2
\frac{p_{\alpha'}}{p_{\alpha}}.
\end{equation}
The transmission amplitude $T_{\alpha \alpha'}$ in Eq. (8) is
equal to $\psi_{\alpha'}(x_N)$ for the problem with the incident
wave in state $\alpha$ and $\varepsilon= E_F$. The
sum runs over all propagating states (for
which $p_{\alpha}$ are real). The generalization of Eq. (8) to
finite temperatures is straightforward, see, e.g., Eq. (7) in Ref. 10.
The actual number $M$ of transverse subbands involved in the
numerical calculations is determined by the condition that further
increase of $M$ does not produce any perceptible change of the
calculated wave functions and transmission coefficients. For the
calculations described below it is sufficient to take $M$ between
10 and 15; this results in a reasonably short calculation time.

\subsection{Numerical results}

In order to decrease the number of the unknown  parameters,
we restricted ourselves to the flat-band approximation, $U(x,y)=0$,
in all calculations.  As a result, we have
only the three geometrical parameters $a$, $h$ and $b$, which
are assumed to be controlled by the gates. The first of them, $a$,
characterizes the width of the constriction in its narrowest parts,
although it is somewhat smaller than the constriction width, cf.
Eqs. (1) and (2). Assuming that the depletion of the 2DEG
by the gates follows a linear law, which is confirmed by
our experimental studies of the conductance quantization in a
single constriction, we can directly associate a change of the
value of $a$ with a change of the gate voltages. As for $h$ and $b$, they
describe mostly the shape of the cavity formed
in the stub region and are related, respectively, to its height and width.
It is convenient to measure all these parameters in units of the cut-off
length $a_0=\hbar \pi/\sqrt{2m E_F}$, which is the width of
the hard-wall quantum wire, when it stops conducting, and is equal to
26.5 nm for $E_F=$8 meV, a value of $E_F$
used in all calculations, and $GaAs$ effective mass $m=0.067~m_e$.

Varying the bottom gate voltage $V_B$, in our model, means simply
changing the parameter $a$ while keeping $h$ and $b$ constant. This
describes the case when the lower (bottom) boundary of the conducting
channel is shifted linearly by  $V_B$, while the
upper boundary remains insensitive to this voltage because of screening
by the electron gas inside the channel. These assumptions are supported
by previous calculations of the potential distribution in  homogeneous,
along $x$ axis, split-gate structures \cite{8}. The calculated
dependence of the conductance on $a/a_0$, in the range of the first
and second plateau, for two values of $h$ and $b=w$ is shown in Fig. 10.
The following qualitative features are evident: for small $h$
the transmission pattern shows narrow, 
almost equally spaced minima of resonance reflections of similar shape:
we call this the "regular" pattern. The number of the minima on the
first plateau (5-7) is consistent with the experimentally observed
number (Fig. 5, solid curve). As $h$ increases ($V_T$ less negative),
the oscillations become irregular and show broad troughs,  
on which are superimposed the closely spaced resonances, and the average
conductance is considerably smaller than that of the shorter cavity.
These results are consistent with experimental observations (Fig. 5, dotted
 curve). Results for a narrower cavity show a similar but less regular pattern.

The situation is more complex when the top gate voltage
$V_T$ is varied. Making $V_T$ less negative obviously leads to an increase
of $h$, and, as mentioned earlier, it also widens
the constriction. To describe this situation 
$a$ and $h$ should be varied. Very likely the cavity width $b$
also changes in this situation, but we expect this change to be small
and make the following assumptions in order to generate the numerical
results: $b$ remains unchanged while $h$ varies linearly as
$h=h_0 + v a$ as $V_T$ is changed. Since the distance of the top
gate from the highest part of the cavity is smaller, but not much
smaller, than its distance from the narrowest part of the constriction,
we expect $v$ to be positive but not much larger than unity.

The dependence of $G$ on $a/a_0$, for fixed $b=w$ and different $h_0$
and $v$, is shown in Fig. 11 for the region of the first 
plateau. The  curves marked (1) and (2) are obtained, respectively,  with
the   $h_0=1.2 a_0$, $v=$1, for (1), and
$h_0=0.6 a_0$, $v=$2 for (2). The parameters are chosen in such a way
that at the beginning of the plateau ($a/a_0 \simeq$ 0.6) the cavity height
is equal to 1.8 $a_0$ for both curves. Curve (1), with smaller $v$, shows
an oscillatory $G$ with regular minima, similar to those
of curve (2) in Fig. 10 for constant $h$. But there is also a difference: the
average conductance shows a small depression near the end of the plateau.
As $v$ increases (curve (2)), the depression becomes more 
pronounced and broad, spreading over the second half of the plateau.
Similar results, but with a less regular oscillatory pattern,
are obtained for cavities with larger $h$, e.g., with 
$h \simeq 2.6 a_0$ at the beginning of the plateau. Also similar qualitative 
features have been obtained for narrower cavities ($b=0.75 w$).

We do not discuss separately the case when the side-gate voltage
$V_S$ is varied, with $h$ and $a$ changing simultaneously, since
we expect the effect to be the same as when $V_T$ changes.
The experimental results of  Figs. 3 and 4 show no qualitative difference 
between the  two cases and confirm this assessment.

\section{DISCUSSION AND CONCLUSIONS}

Comparison of the experimental and theoretical results shows that the
qualitative behavior of the ballistic conductance, as a function of
different gate voltages, is the same in either case. This indicates that
our choice of parameters and the assumptions about their variation with
gate voltages, supported in part by the solution of the electrostatic
problem, represent fairly the experimental situation. Below we discuss
this in more detail.

When we make $V_S$ and $V_T$ more negative than $V_B$, the conducting
channel is shifted towards the bottom gate and the cavity height
$h$ decreases. In this situation we obtain rather regular oscillations
of the conductance as a function of bottom, top or side gate voltages.
The minima correspond to resonant reflections of the electron
waves from quasibound states in the cavity \cite{11}. Variation
of one of the gate voltages sweeps the levels of the quasibound
states through the Fermi level $E_F$ and results in a resonance minimum
each time $E_F$ coincides with one of them.
The regularity in shape and spacing of these minima, for cavities with
small heights, follows from the fact that these minima originate from
the same set of quasibound states. For confirmation, we
calculated again curves (1) in Figs. 10 and 11 with just two
{\em transverse} cavity modes in the expansion (4); in the region of the first
plateau the first mode is transmitted through the constrictions,
while the second one is quasibound in the cavity. Comparing these
results with the curves (1) of Figs. 10 and 11, we find that all
regular resonances occuring on the first plateau appear also in this
simplified calculation and their shapes are very similar. We,
therefore, conclude that the regular resonance reflection pattern
results from quasibound states associated with the second {\it transverse}
mode and  each of them characterized by a proper
wavenumber that expresses {\em longitudinal} quantization
along the $x$ direction.
Higher transverse quantization states bring additional resonant
features and make the dependence of the conductance on the
gate voltages less regular. The experimental results of Figs. 3-5 (solid
curves) show  well-defined regularly spaced minima in $G$ and fully
corroborate this analysis. Note that the number of the minima obtained
from the theory is about the same as that observed experimentally; this means
that the cavity width is really close to the lithographic width $2w$ of
the stub, as we assumed in our model. The experimental dependences of the
conductance on the top, side, and bottom gate voltages are similar and
show almost the same number of oscillations. This is in agreement
with our assumption that we are dealing with a wide cavity of small height.
For the same $h$ the  minima are narrower for
wider cavities since in such a case the upper boundary of the cavity
is smoother and the electron motion through the cavity is more adiabatic.
In terms of our model this means that the derivative $d y_{cav}(x)/ d x$
is smaller. Experimentally, a broadening of the cavity without
changing its height can be achieved by applying less negative
voltages to the side gates. 

When we make $V_S$ and/or $V_T$ less negative 
compared to the values of the "regular" case discussed above,
the channel widens and the cavity height $h$ increases.
In this "irregular" case the situation is modified as follows.
i) More quasibound states occur and influence the resonant reflection,
ii) The coupling between transmitted and quasibound states increases
due to the increase of the non-adiabaticity. Both these factors
should lead to a decrease of the average conductance $G$.
This decrease of $G$  is clearly seen in the theoretical results, cf.
Figs. 10-11. Besides, the theory shows that the behavior of $G$  
as a function of different gate voltages has a qualitative difference:
the variation of $V_B$ produces broad minima and depression of the average
$G$ over the entire plateaus while the variation of $V_T$ or $V_S$
gives narrower minima and the average $G$
has pronounced troughs near the ends of the plateaus. Our theoretical
calculations show that the troughs are more pronounced when the
height $h$ is larger and when it increases faster with
the opening of the constriction. A simple explanation is as
follows. Sweeping $V_T$ or $V_S$ towards less negative values
along the conductance plateau substantially increases
$h$ and leads to a departure from the short-cavity, "regular" pattern
to the "irregular" one of a long-cavity. Therefore, the averaged $G$  
decreases at the end of the first plateau and then increases, when
the second transverse mode is allowed in the constriction, reaching
the next plateau. If the initial value of $h$ is 
already large, we need a lesser increase in $h$ in order to move to
the long-cavity regime; then the decrease of the average $G$  
appears earlier and gives rise to a broader trough. The experimentally
observed behavior of $G$, cf. dashed curves in Figs. 3-5,
shows pronounced troughs and a decrease in its average value
in agreement with the above interpretation.
We emphasize that the appearance of these troughs in $G$
is a very common and reproducible feature of the stubbed
quantum devices studied  in this work.

Possible EST applications require a regular, periodic dependence
of the conductance on the gate voltages. The initial idea$^1$, followed
in subsequent theoretical works, was to use a narrow cavity, containing
{\it one} quantized state, or mode,
in the ($x$) transport direction, and control the conductance through it
by changing the height of the cavity by a top-gate voltage.
However, from our results it is clear that real EST devices
do not satisfy these conditions because the electrostatic potential
created by the gates is rounded near the corners, cf.  Fig. 8. 
Moreover, for a real device, as this work shows, if the cavity is long enough,
the dependence of the conductance on the gate voltages is rather
irregular. We have shown that it
is possible to obtain a short-length cavity by a proper choice  
of the gate voltages that is wide in the transport direction.
The conductance through such a cavity shows a regular pattern  of
resonant reflection minima associated with the  quasibound
states in it that result from {\em longitudinal} quantization.

\acknowledgements

One of us (P. D.) gratefully acknowledges the award of a Senior Research 
Associateship by the National Research Council, Washington, DC.
The work of P. V. was supported by the Canadian NSERC Grant No. OGP0121756.

\begin{figure}
\caption{Schematic drawing of an EST device. Lithographic dimensions:
$L=0.30 \mu$m, $2W=0.25 \mu$m. S1, T, S2, and B are surface Schottky gates.
The area enclosed between the dashed lines defines the conducting channel.}
\label{fig. 1}
\
\caption{Scanning electron micrograph of a fabricated EST device. Lighter
areas indicate the Schottky gates.}
\label{fig. 2}
\
\caption{Conductance as a function of the top gate voltage $V_T$
with $V_{S1}=-860$ mV and $V_{S2}=-964$ mV.
The solid and dotted curves correspond
to $V_B=-765$ mV and $V_B=-780$ mV, respectively. The arrows indicate
locations of broad depressions in $G$.}
\label{fig. 3}
\
\caption{Conductance as a function of the side gate voltage $V_S$
 ($V_S=V_{S1}=V_{S2}+104$ mV)
at fixed $V_B$ and $V_T$.
The solid curve was obtained at $V_T=-1400$ mV and $V_B=-755$ mV. The dotted 
curve  corresponds
to $V_T=-800$ mV and $V_B=-790$ mV. The arrow shows the position of a 
broad depression in $G$.}
\label{fig. 4}
\
\caption{Conductance as a function of the bottom gate voltage $V_B$
with $V_{S1}=-860$ mV and $V_{S2}=-964$ mV at fixed $V_T$.
The solid and dotted curves correspond
to $V_T=-1085$ mV and $V_T=-635$ mV, respectively. 
The arrows indicate  locations of  
broad depressions in $G$.}
\label{fig. 5}
\
\caption{Conductance as a function of the top gate voltage $V_T$
with $V_{S1}=-860$ mV, $V_{S2}=-964$ mV, and $V_B=-765$ mV at different
temperatures.}
\label{fig. 6}
\
\caption{Conductance as a function of the top gate voltage $V_T$
with $V_{S1}=-860$ mV, $V_{S2}=-964$ mV .
$V_B=-765$ mV for 
different source-drain voltages.}
\label{fig. 7}
\
\caption{ Contour plot of $\phi(x,y,z)$ created in the plane of the 2DEG,
at $z=d=$80 nm, of a two-gate EST for (a) $V_B=V_T$ and (b) $V_B-V_T=$0.2 V.
The interval between equipotential lines is 
6.6 mV  in (a)  and 8.4 mV. The thick solid lines show the edges of the gates.}
\label{fig. 8}
\
\caption{ Model of the conducting channel for a four-gate EST  
(Figs. 1 and 2). The area enclosed between the traces $y_t$ and $y_b$
defines the shape and size of the conducting channel under gate biases that was
used in our calculations. The solid straight lines are the lithographic 
 edges of the Schottky gates S1, T, S2, and B.}
\label{fig. 9}
\
\caption{ Conductance as function of $a/a_0$
for two different heights of the stub cavity when its width
is equal to the lithographic width ($b=w$): (1) $h=2a_0$; 
(2) (shifted for clarity) $h=3a_0$. This corresponds to the experimental
condition when  $V_B$ is varied with $V_T$ and $V_S$ constant. }
\label{fig. 10}
\ 

\caption{Conductance as function of $a/a_0$ for $b=w$ and $h=h_0+va$.
(1) $h_0=1.2 a_0$, $v=$1; (2)
$h_0=0.6 a_0$, $v=$2. This corresponds to the experimental
condition when  $V_T$ is varied with $V_B$ and $V_S$ constant.}
\label{fig. 11}

\end{figure}
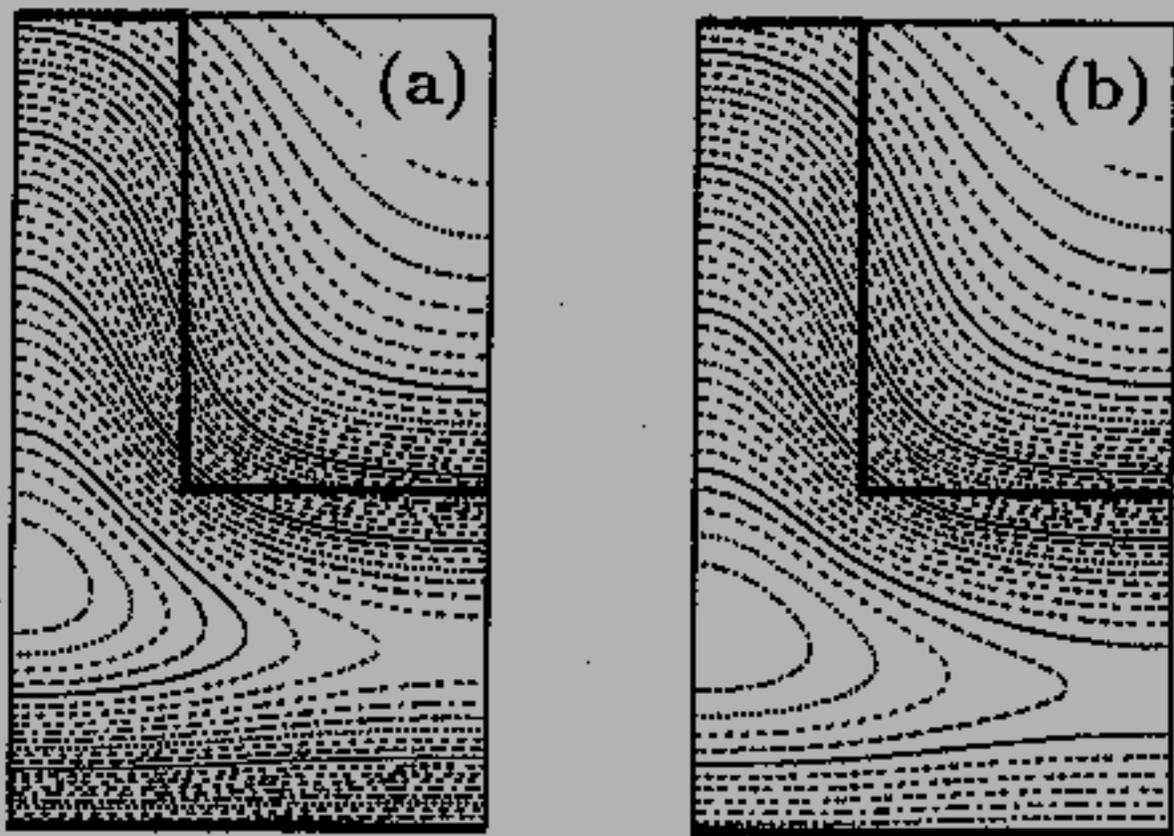
\end{document}